\documentclass[journal,comsoc]{IEEEtran}
\usepackage[T1]{fontenc}
\usepackage{easyReview}
\usepackage{amsmath}

\usepackage{comment}

\usepackage{multirow}
\usepackage{comment}
\usepackage{hyperref}
\usepackage{cite}
\usepackage{dblfloatfix}
\usepackage{graphicx}
\usepackage{amsmath}
\usepackage{cleveref}
\usepackage{caption}
\usepackage[caption=false,font=footnotesize]{subfig} 
\usepackage{lipsum}
\usepackage{float}

\crefname{figure}{fig.}{figs.}           

\begin{document}


\title{DV-QKD Coexistence With 1.6 Tbps Classical Channels Over Hollow~Core Fibre}

%

\author{
    Obada~Alia, Rodrigo~S~Tessinari, Sima~Bahrani, 
    Thomas~D~Bradley,
    Hesham~Sakr,
    Kerrianne~Harrington,
    John~Hayes,
    Yong~Chen,
    Periklis~Petropoulos,
    David~Richardson,
    Francesco~Poletti,\\
    George~T~Kanellos,
    Reja~Nejabati, Dimitra~Simeonidou

\thanks{O.~Alia, S.~Bahrani, G.~Kanellos, R.~Nejabati and D.~Simeonidou are with the High Performance Network Group School of Computer Science, Electrical \& Electronic Engineering and EngineeringMaths (SCEEM), University of Bristol BS8 1TH, Bristol, UK, e-mail: (obada.alia@bristol.ac.uk).}
\thanks{R.~S.~Tessinari is with Toshiba Europe Limited, Cambridge Research Laboratory, 208 Science Park, Milton Road, Cambridge CB4 0GZ, UK. Previously with the High Performance Network Group at the University of Bristol}
\thanks{T.~D.~Bradley is with the High-Capacity Optical Transmission Laboratory, Eindhoven University of Technology, 5600 MB, Eindhoven, The Netherlands.}
\thanks{H.~Sakr, K.~Harrington, J.~Hayes, Y.~Chen, P.~Petropoulos, D.~Richardson and F.~Poletti are with the Optoelectronics Research Centre, University of Southampton, Southampton SO17 1BJ, UK.}
}

\markboth{\alert{Journal of Lightwave Technology, VOL. XX, No. X, 2022.}}{}

\maketitle

\begin{abstract}

The feasibility of coexisting a quantum channel with carrier-grade classical optical channels over Hollow Core Nested Antiresonant Nodeless Fibre (HC-NANF) is experimentally explored for the first time in terms of achievable quantum bit error rate (QBER), secret key rate (SKR) as well as classical signal bit error rates (BER). A coexistence transmission  of 1.6 Tbps is achieved for the classical channels simultaneously with a quantum channel over a 2 km-long HC-NANF with a total coexistence power of 0 dBm. To find the best and worst wavelength position for the classical channels, we simulated different classical channels bands with different spacing between the quantum and classical channels considering the crosstalk generated from both Raman scattering and four-wave-mixing (FWM) on the quantum channel. Following our simulation, we numerically estimate the best (Raman spectrum dip) and worst locations (Raman spectrum peak) of the classical channel with respect to its impact on the performance on the quantum channel in terms of SKR and QBER. We further implemented a testbed to experimentally test both single mode fibre (SMF) and HC-NANF in the best and worst-case scenarios. In the best-case scenario, the spacing between quantum and classical is 200~GHz (1.6~nm) with 50~GHz (0.4~nm) spacing between each classical channel. The SKR was preserved without any noticeable changes when coexisting the quantum channel with eight classical channels at 0~dBm total coexistence power in HC-NANF compared to a significant drop of 73\% when using SMF at $-24$~dBm total coexistence power which is 250 times lower than the power used in HC-NANF. In the worst-case scenario using the same powers, and with 1~THz (8~nm) spacing between quantum and classical channels, the SKR dropped 10\% using the HC-NANF, whereas in the SMF the SKR plummeted to zero.
\end{abstract}


%
\IEEEpeerreviewmaketitle

\section{Introduction}
\label{sec:intro}

Quantum communication networks present a revolutionary step in the field of quantum communication \cite{pirandola2020advances}. In the last two decades, quantum networks evolved from relayed point-to-point quantum communication links - with main focus in quantum security applications such as QKD - to multiuser quantum networks with partial or full connectivity beyond point-to-point links functioning in real life environment \cite{mehic2020quantum}. However, one of the main goals of quantum communication is to provide a worldwide connectivity via complex networks – similar to the current internet – with an ultimate security based on the laws of physics rather than computational complexity \cite{RevModPhys.81.1301}. The Quantum Internet vision is moving towards interconnecting seamlessly multiple nodes and enabling applications beyond QKD such as blind and distributed quantum computing \cite{van2016path}. For such applications to reach their maximum capabilities, a seamless medium allowing the coexistence of high power classical channels with quantum channels would be beneficial for the integration of quantum technologies with the current optical infrastructure.

QKD is an implementation of quantum cryptography technologies that utilises the basic principles of quantum mechanics such as the uncertainty principles to generate information secure symmetric keys that could be used to encrypt the decrypt key information \cite{pirandola2020advances}. QKD has been considered as the ultimate physical layer security since the quantum-grade encryption mechanism proves to be information theoretical secure \cite{gisin2002quantum}. In discrete variable QKD (DV-QKD) \cite{gisin2002quantum}, the process of generating the symmetric key is undertaken by transmitting encoded single photons between the sender and receiver, Alice and Bob respectively over an authenticated quantum channel. This quantum channel is also supported by classical communication channels for basis reconciliation, error correction and privacy amplification protocols \cite{elkouss2010information}. 


The integration of quantum technologies with the current classical optical networking infrastructure would help in the wider deployment of QKD schemes. However, since the commercially deployed DV-QKD schemes rely on the exchange of single or few photons between Alice and Bob and are limited by the power budget of the optical link used for quantum channel, deployment of quantum channels coexisting with classical optical communication channels is therefore a very challenging task. To further elaborate on the challenges of such schemes, the optical power of classical communication channels is orders of magnitude higher than the power of the quantum channel for quantum communication and the additional noise generated from such channels due to optical non-linear effects like Raman scattering and four wave mixing (FWM) \cite{chraplyvy1990limitations} degrades the quantum channel performance extensively and was studied in \cite{eraerds2010quantum, patel2012coexistence, zavitsanos2019coexistence,da2014impact, subacius2005backscattering}. Therefore, coping with optical nonlinearity represents a major challenge for QKD systems. A high isolation (>110~dB) \cite{peters2009dense} between the quantum and classical channel is required and usually is implemented by sharp filters or cascading multiple filters to partly isolate the noise from the classical channels and distinguish the quantum signal from adjacent channel crosstalk; hence improving the quantum signal. Nevertheless, non-linear processes in glass fibres also generate excessive photons at the same wavelength as the quantum channel (in-band noise) that cannot be mitigated or filtered, dramatically affecting its performance. 






The coexistence of quantum and classical channels where they are both in the same optical wavelength bands (C-band) or in different optical wavelength bands (O-band for the quantum and C-band for the classical) has been extensively studied and demonstrated in \cite{patel2012coexistence, chapuran2009optical, eraerds2010quantum, dynes2016ultra, hugues202011}. Coexisting the quantum and classical channels in the C-band has many advantages, such as lower fibre loss in the C-band comparing to the O-band and better compatibility with the current fibre infrastructure. However, the nonlinear effects such as Raman scattaring becomes more dominant, hence a lower total coexistence power is achievable as recorded from previous implementations, namely $-18.5$~dBm, $-22$~dBm and $-25.5$~dBm in \cite{patel2012coexistence,eraerds2010quantum,dynes2016ultra} respectively. In contrast, when coexisting the quantum and classical channels in different bands, the total coexistence power could be as high as $-2$~dBm as shown in \cite{chapuran2009optical}.

Here, we consider a radical solution to this challenge by adopting a transmission fibre technology, in which nonlinearities are simply not present. Antiresonant Hollow Core Fibres (AR-HCF) are optical fibres where light is guided in a hollow core via anti-resonance effects within the glass membranes surrounding the core \cite{gao2018hollow}. AR-HCF surpass the performance of glass core fibres as they have ultra-low optical mode overlap with the glass (and hence reduced optical nonlinearity and Rayleigh scattering), a lower latency, a very low chromatic dispersion, an unprecedented polarisation purity and ultra-low back scattering \cite{sakr2020hollow, taranta2020exceptional, michaud2021backscattering}. Significant improvement in optical performance of HCFs were made possible by inducing new features such as negative curvature core surround \cite{wang2011low}, then by fibre topologies with a single layer of non-touching tubes \cite{kolyadin2013light}, and finally by the addition of small nested tubes - a design known as hollow core Nested Antiresonant Nodeless Fibre (HC-NANFs) \cite{poletti2014nested}. HC-NANFs provide a solution for coexistence scheme as they do not only provide several attractive advantages compared to glass core fibres, such as reduced optical nonlinearity and ultra-lower latency, but also have the ability to reach a total loss value lower than that of conventional solid single-mode fibres \cite{poletti2014nested}. These desirable qualities enable the transmission of classical channels at high optical powers while coexisting with quantum channels over the same medium \cite{alia20211}. The losses of HC-NANF have improved immensely from 1.3~dB/km \cite{bradley2018record} (used in this experiment), to 0.28~dB/km \cite{jasion2020hollow} and recently to 0.22~dB/km \cite{sakr2021hollow}. 




In this paper, we demonstrate the coexistence of a 16 dB power budget commercial DV-QKD system (Clavis3 \cite{CL3}) and 8~x~200~Gbps 16-QAM carrier-grade classical optical channels at an extremely high total coexistence power of 0~dBm over a 2~km HC-NANF, revealing minimal effects on the quantum channel performance. We also numerically estimate  the best and worst wavelength location for coexisting an 8-channels classical band with a fixed quantum channel and compare the QKD performance in terms of SKR and QBER for both scenarios in HC-NANF and SMF. The paper is organized as follows: in Section \ref{sec:testbed} we discuss the experimental system setup followed by Section \ref{sec:simulation} where we discuss the numerical analysis for the classical channels position. Section \ref{sec:result} present the results and is divided into three parts, optical spectrum for quantum/classical coexistence \ref{sec:sepctrum}, quantum system characterisation \ref{sec:Characterisation} and coexistence analysis based on the quantum channel performance \ref{sec:Co-propagation_results}. Finally, in Section \ref{sec:conclusion} we conclude the paper.

\section{Experimental System Setup}

\label{sec:testbed}
\begin{figure*}[h]
    \centering
    \includegraphics[width=1\linewidth,clip]{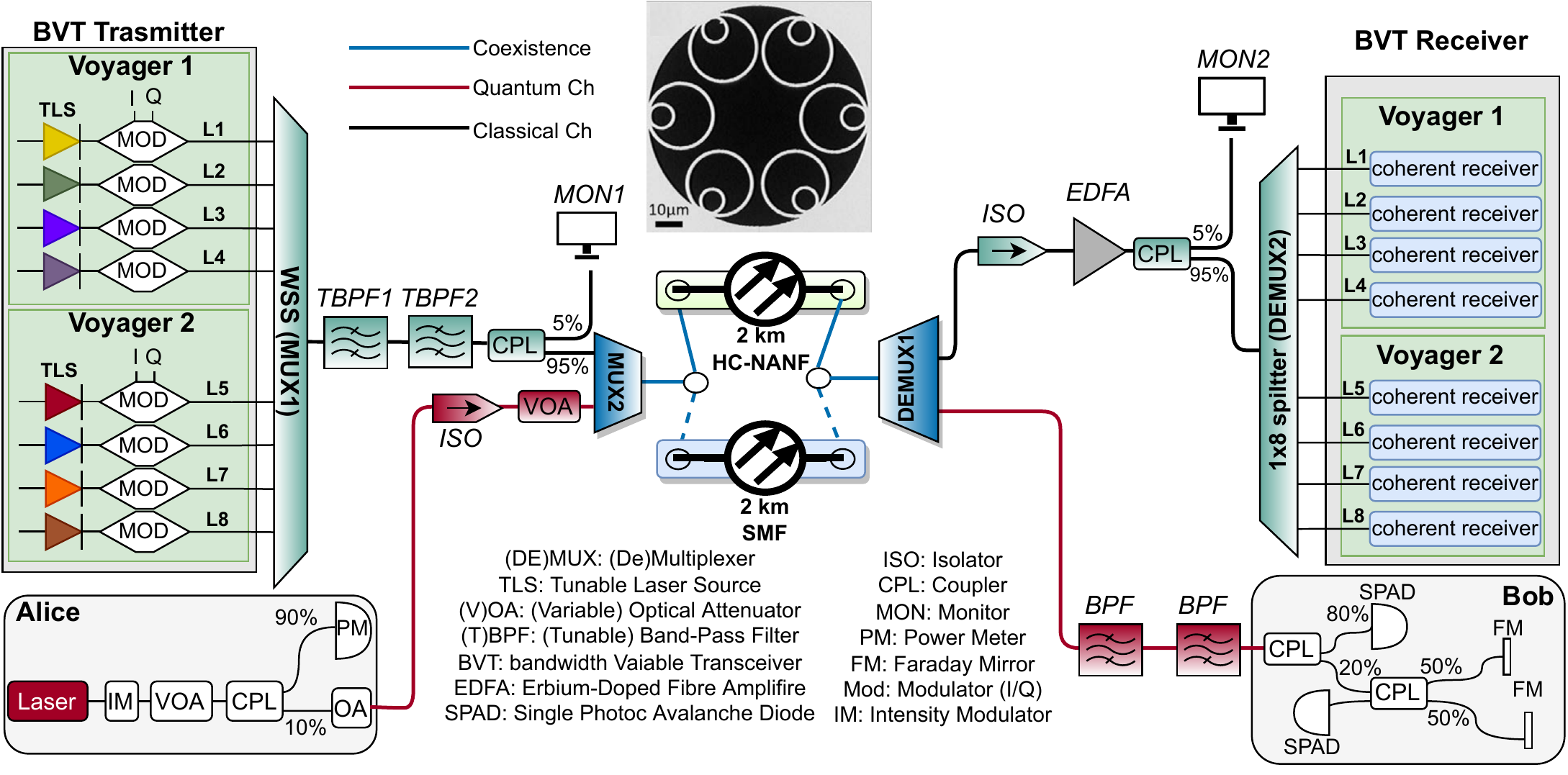}
    \caption{Experimental Testbed for the Coexistence of 1.6 Tbps classical channels and DV-QKD channel over 2km HC-NANF and SMF. Inset: scanning electron micrograph (SEM) of the  HC-NANF cross section.}
    \label{fig:testbed}
\end{figure*}

\Cref{fig:testbed} shows the experimental system setup used to demonstrate the coexistence of eight classical channels with a DV-QKD channel over a 2 km-long HC-NANF and SMF (as inset figure, \Cref{fig:testbed} also illustrates the cross-sectional diagram of the HC-NANF used \cite{bradley2018record}). The testbed facilitates the experimental evaluation of the nonlinear effects on the performance of the quantum channel created by the presence of classical channels spectrally close to the quantum channel using both HC-NANF and SMF. For the classical channels, two optical packet DWDM platforms are used with bandwidth-variable transponders (BVTs) \cite{voyager}. Each of these units include four BVT ports reconfigurable to coherent 100~Gbps (PM-QPSK), 150~Gbps (8-QAM) or 200~Gbps (16-QAM) and each port can be tuned to any of the 100 wavelengths in the C-band included in the ITU-T grid with 50~GHz offset. Adaptable soft-decision forward error correction (SD-FEC) is also available in the transponders to enable maximum transmission capacity with minimum errors. Since the HC-NANF has ultra-low optical nonlinearity, it allows the transmission of classical channels at an extremely high power (over 0~dBm) with minimal effect on the quantum channel. Therefore for the HC-NANF since we are using high total coexistence power, all the ports were configured with the 16-QAM modulation for a maximum capacity of 200~Gbps per channel resulting in 1.6~Tbps of transmission overall and the SD-FEC considered was 15\% with a detector sensitivity of -26~dBm. As for the SMF, due to the nonlinear effects in this type of fibre the coexistence power was reduced compared to the HC-NANF case, and a substantially higher detector sensitivity is required to receive an error-free classical signal. Therefore, for the SMF a PM-QPSK modulation was used for a maximum capacity of 100~Gbps per channel resulting in 800~Gbs of transmission overall and the SD-FEC considered was 25\% with a detector sensitivity of -35~dBm. \\
For the quantum channel, IDQuantique DV-QKD systems are used (Clavis3 QKD Platform \cite{CL3}). These systems are implemented to run with the COW (Coherent One-Way) protocol \cite{stucki2009continuous}. The aim of the COW protocol is to make the implementation of a QKD system as simple as possible using only two single photon detectors and to allow a significant increase of the final secret key rate. Moreover in the Clavis3 system, Alice and Bob encode bits with quantum state carried by single photons (qubits) using time-bin encoding method by creating a pair of coherent pulses propagating in the same spatial mode and separated by a given time. The first pulse is called the early pulse and the second one is called the late pulse. Hence, the measurement method to analyse this basis is simply to measure the time of detection of the optical pulse. If one detection occurs in the early time-bin, the qubit value is a $|0\rangle$ state, whereas if it occurs in the late time-bin, the qubit is a $|1\rangle$ state.

\begin{table}[h]
\centering
\caption{Parameters for HC-NANF Coexistence Testbed}
\vspace{-0.3cm}
\label{tab:parameters}
\begin{tabular}{cc}
\hline
\hline
\multicolumn{1}{c}{Parameters}  &   \multicolumn{1}{c}{Value}       \\
\hline
\multicolumn{2}{c}{\textit{Classical Channels}} \\
Number of Channels  &   8   \\


\begin{tabular}[c]{@{}l@{}} \ \ \ Classical Channel\\ Frequencies scenario 1\end{tabular}         & \begin{tabular}[c]{@{}l@{}@{}l@{}}193.50 THz, 193.45 THz,\\ 193.40 THz, 193.35 THz, \\193.30 THz, 193.25 THz,\\ 193.20 THz, 193.15 THz\end{tabular} \\ 

\\  

\begin{tabular}[c]{@{}l@{}} \ \ \ Classical Channel\\ Frequencies scenario 2\end{tabular}         & \begin{tabular}[c]{@{}l@{}@{}l@{}}192.70 THz, 192.65 THz,\\ 192.60 THz, 192.55 THz, \\192.50 THz, 192.45 THz,\\ 192.40 THz, 192.35 THz\end{tabular} \\

Grid Spacing        &   50 GHz      \\
Modulation Format   &   16-QAM     \\
\begin{tabular}[c]{@{}l@{}}Optical Signal-to-Noise\\  \ \ \ \ \ Ratio (OSNR)\end{tabular} & 20 dB \\
Transmission rate per channel & 200 Gbps      \\
Total transmission rate      &   1.6 Tbps    \\
Pre-FEC Level       &   15\%        \\
Detector sensitivity*& -26 dBm      \\
                                    \\
\multicolumn{2}{c}{\textit{Quantum Channel}}    \\
DV-QKD Wavelength   & 1547.72 nm     \\
DV-QKD Frequency   & 193.70 THz     \\
QKD Protocol        & COW          \\
Maximum Loss    & 16 dB \\
                                    \\
\multicolumn{2}{c}{\textit{Optical Band Pass/Rejection Filter (OBRF)}}   \\
\begin{tabular}[c]{@{}l@{}}Insertion loss band pass port\end{tabular}     & 0.5 dB    \\
\begin{tabular}[c]{@{}l@{}}Center wavelength\\ \ \ \ band pass port\end{tabular}   & 1547.72 nm \\
Bandwidth band pass port & 100 GHz   \\
\hline
\hline
\multicolumn{2}{l}{\textit{*Corresponding to 16-QAM Modulation @200 Gbps and back-to-back.}}  \\
\end{tabular}
\end{table}

As shown in \Cref{fig:testbed}, the eight coherent output ports of the two BVTs are multiplexed using a wavelength selective switch (WSS) shown as MUX1 with 5~dB of insertion loss for a total throughput of 1.6~Tbps. The WSS is used as a multiplexer and a band pass filter to couple the classical channels into a single fibre and provide a 30~dB isolation. The WSS combined output is connected to the input port of a tunable band pass filter (TBPF1) with 60~dB of isolation and extremely sharp filter edges followed by a Gaussian shape tunable band pass filter (TBPF2) with 30~dB of isolation to further suppress the noise generated by the classical channels. The classical channels are then coupled to the quantum channel through a WDM multiplexer (DWDM add drop filter) MUX2 in a co-propagation coexisting configuration and travels through a 2~km of HC-NANF or SMF. After the coexistence within the selected fibre, the output is connected into a WDM demultiplexer tuned to 1547.72~nm (DEMUX1) that passes the quantum channel while rejecting all other channels (classical channels). After the WDM demultiplexer, the quantum signal is passed through a double stage filtering using two fixed WDM band-pass filters centred at the quantum channel wavelength (1547.72~nm) to eliminate any noise generated by the classical channels. The output of the second filter is connected to the DV-QKD Bob unit, which undertakes the single-photon detection and processing of the encoded photons allowing the completion of the COW protocol of the DV-QKD systems. The output of the reflection port of the WDM demultiplexer is directed to an optical isolator with insertion loss of 1~dB to prevent the tunable laser used by the BVT Rx as a local oscillator from returning to the Bob-QKD unit and interfering with the QKD measurements. It also prevents the Amplified Spontaneous Emission (ASE) noise generated by the Erbium-Doped Fibre Amplifier (EDFA) which is used to set the classical signals to the suitable detectable power levels before demultiplexing via a 1x8 splitter. Furthermore, an optical isolator is used after the Alice-QKD unit to prevent the noise generated from the classical channels returning to the Alice-QKD unit and interfering with the quantum signal. After the optical isolator at the Alice side, a variable optical attenuator (VOA) is used to adjust the losses of the quantum channel to the minimum operational level of 10~dB to optimise the functionality of the Clavis3 QKD system and to prevent over saturating the single photon detector in the Bob-QKD unit. It also allows us to obtain a direct comparison of the quantum/classical coexistence between the SMF and HC-NANF in terms of losses. In this testbed, the total end-to-end (Alice to Bob) quantum channel loss is 10.5~dB. A summary of main parameters of the testbed is shown in Table \ref{tab:parameters}.

\section{Channel Spacing/Position for Quantum/Classical Coexistence }
\label{sec:simulation}

For the classical channels, we consider a classical band of 8~channels and 50~GHz spacing between each channel, coexisting with a fixed quantum channel (193.7~THz) in SMF. In \cite{bahrani2018wavelength,ou2018field,bahrani2019resource}, optimal wavelength assignment in hybrid quantum-classical systems have been investigated. Our calculation is based on the fact that the classical channels are located in the C-band and the quantum channel is at ITU channel 37 (193.7~THz). Furthermore, it is assumed that the quantum channel is in the anti-stokes region of the Raman spectrum of the classical channels. We denote the coexistence power by $I_{\rm c}$. Considering both Raman scattering and four-wave-mixing (FWM), we numerically calculate the best and worst locations of a classical band with respect to its impact on the performance of the quantum channel in terms of SKR and QBER. \Cref{fig:SMF_classical_channels} shows the best and worst cases for the spacing between the classical band and the quantum channel for different values of $I_{\rm c}$. For low values of $I_{\rm c}$, Raman scattering is the dominant source of noise. Therefore, the optimum location of the classical band is at the dip of the Raman spectrum which is near the quantum channel (100-200~GHz spacing), whereas the worst case is at the peak of the Raman spectrum (1.0-1.4~THz spacing). As $I_{\rm c}$ increases, both Raman noise and FWM crosstalk increase as well, however, FWM crosstalk grows much faster and becomes more dominant at higher coexistence power $\approx$	-7~dBm. Since FWM products close to the classical band are stronger, for higher values of $I_{\rm c}$ the space between the quantum channel and the classical band should be chosen such that the FWM products with high power $f_{ijk} = f_i \pm f_j \pm f_k$ do not lie in the quantum channel filter bandwidth \cite{he2014four}. In \Cref{fig:SMF_classical_channels}, the photon counts for FWM noise and Raman noise in the special case of 200~GHz (1.6~nm) spacing are also depicted. Here, the time duration of the detector and the detection efficiency are assumed to be $100~{\rm ps}$ and $0.7$, respectively. 


\begin{figure}[h]
    \centering
    \includegraphics[width=1\linewidth,clip]{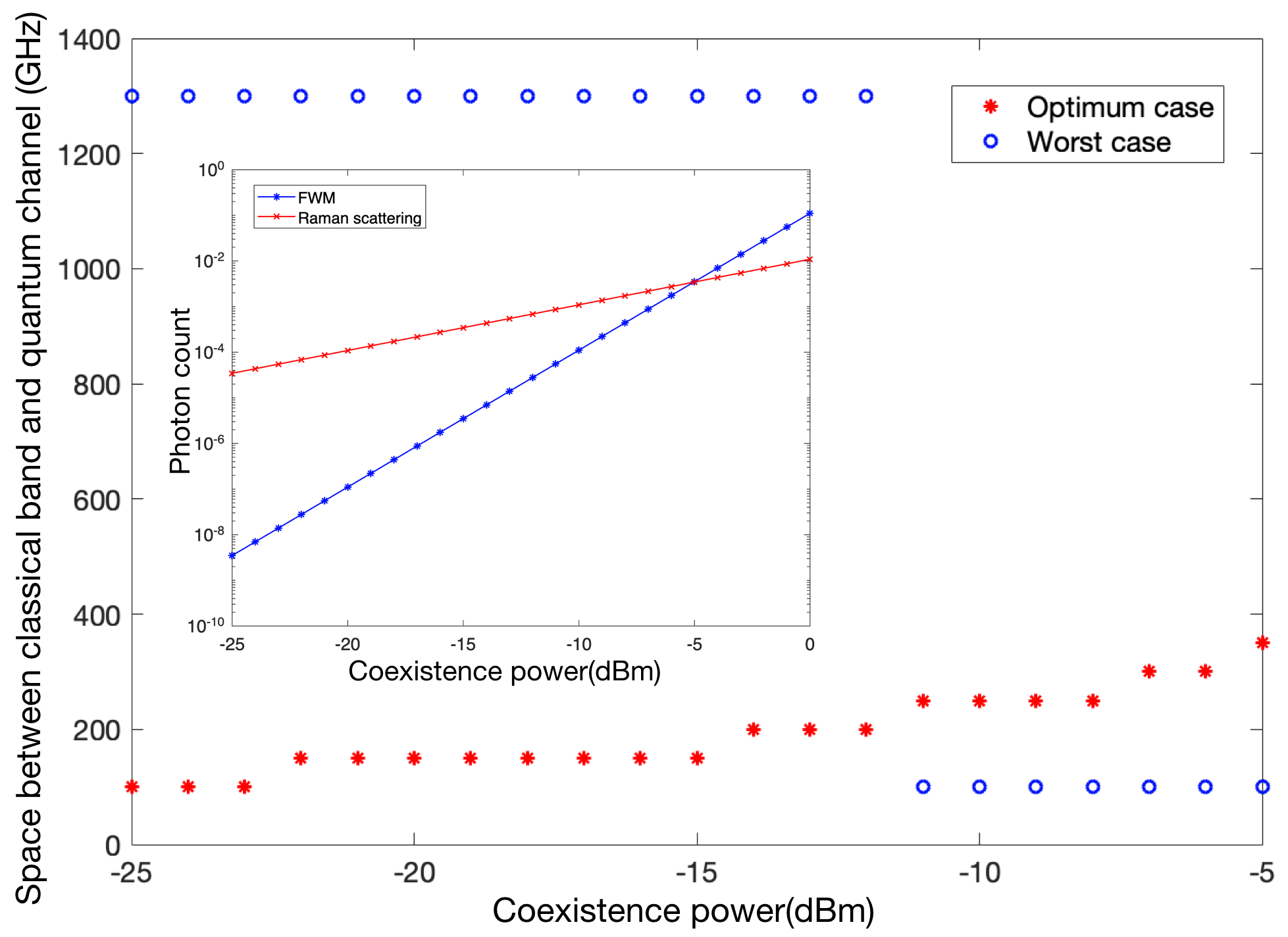}
    \caption{Best and worst case scenarios for the classical band spacing, considering different values for the coexistence power in the SMF. The photon counts for Raman noise and FWM noise in the special case of classical band spacing equal to 200~GHz (1.6~nm) are also shown in the figure.}
    \label{fig:SMF_classical_channels}
\end{figure}

\section{Experimental Results}
\label{sec:result}
\subsection{Optical Spectrum of Quantum and Classical Channels}
\label{sec:sepctrum}
\Cref{fig:spectrum} shows the spectrum of $8~\times~200$ Gbps classical channels coexisting with a quantum channel obtained by using a tap coupler as shown in \Cref{fig:testbed} \textit{MON1 \& MON2} for the best and worst case scenarios. \Cref{fig:spectrum} also demonstrates the optical filters profile (pink) of the tunable band pass filters \textit{(TBPF1 \& TBPF2)} used to filter out the noise generated by the classical channel. It also shows the optical filter profile of the Band-pass filter \textit{(BPF)} used before the Bob-QKD unit (red) to suppress the nonlinear effects from passing to the QKD detectors. The filter profiles were generated by using the output of an EDFA as a noise source. Furthermore, the spectrum of the eight classical channels after the filtering (black) MON1 and the amplification stage (blue) MON2 is shown before demultiplexing and coherent detection at the BVT receiver.

\begin{figure}[h]
    \centering
    \includegraphics[width=0.9\linewidth,clip]{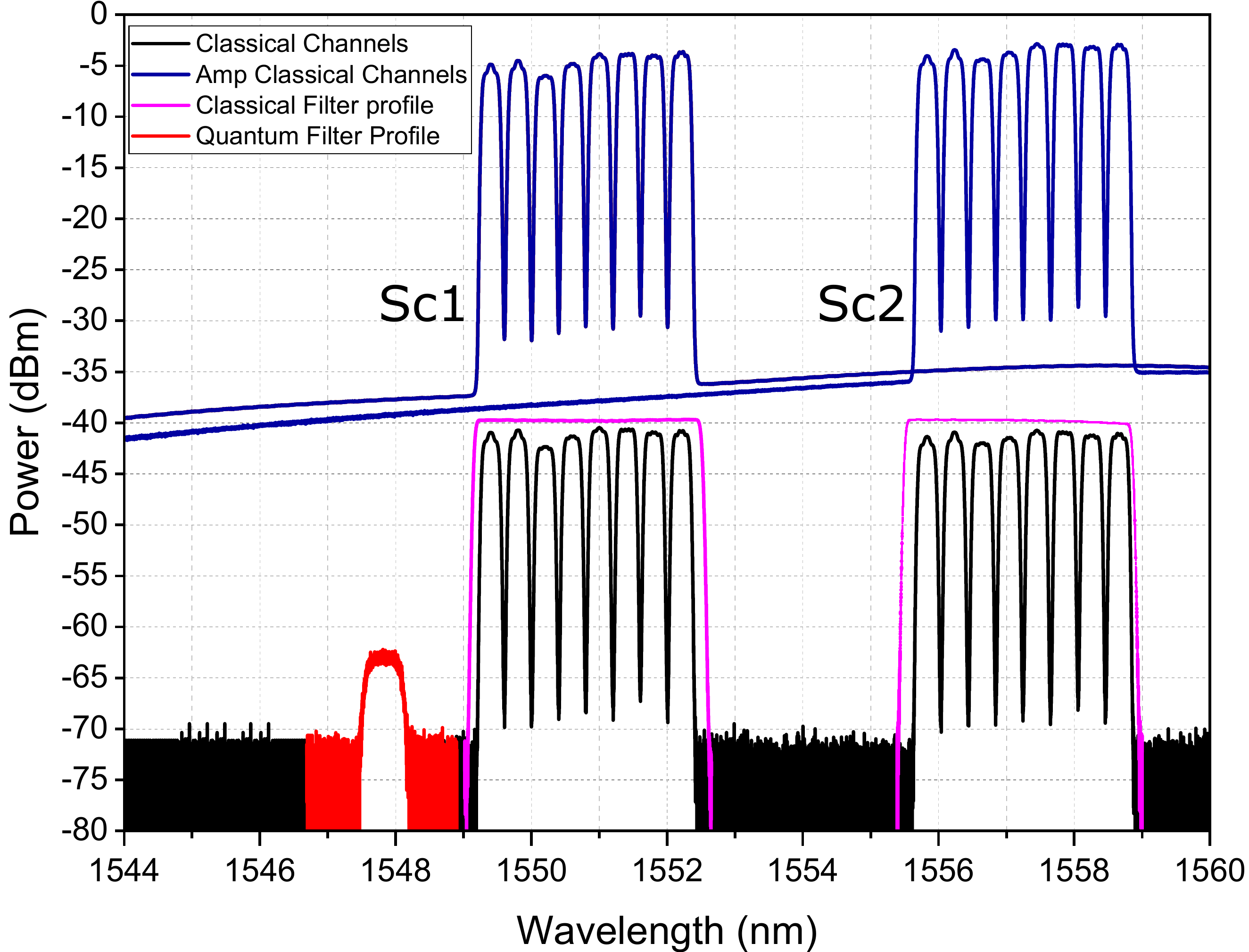}
    \caption{Spectrum of the combined transmission of quantum and classical channels with optical filter profiles of both scenarios.}
    \label{fig:spectrum}
    \setcounter{figure}{4}
\end{figure}

{\newcommand\figureSize{0.475}
\begin{figure*}[b]
    \centering
    \begin{tabular}{cc}
        \subfloat { \includegraphics[width=\figureSize\linewidth]{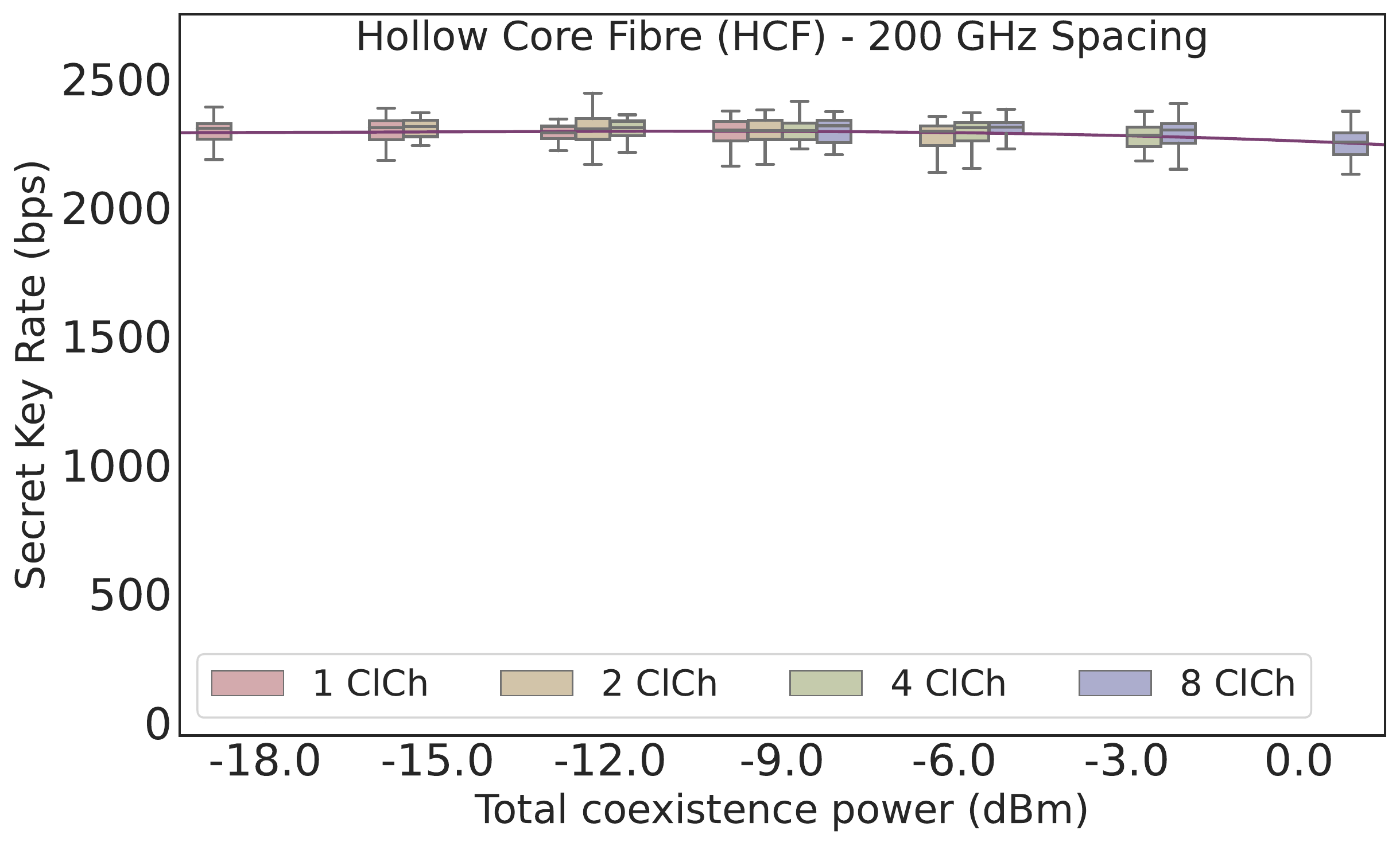}             \label{hcf_200_skr} } & 
        \subfloat { \includegraphics[width=\figureSize\linewidth]{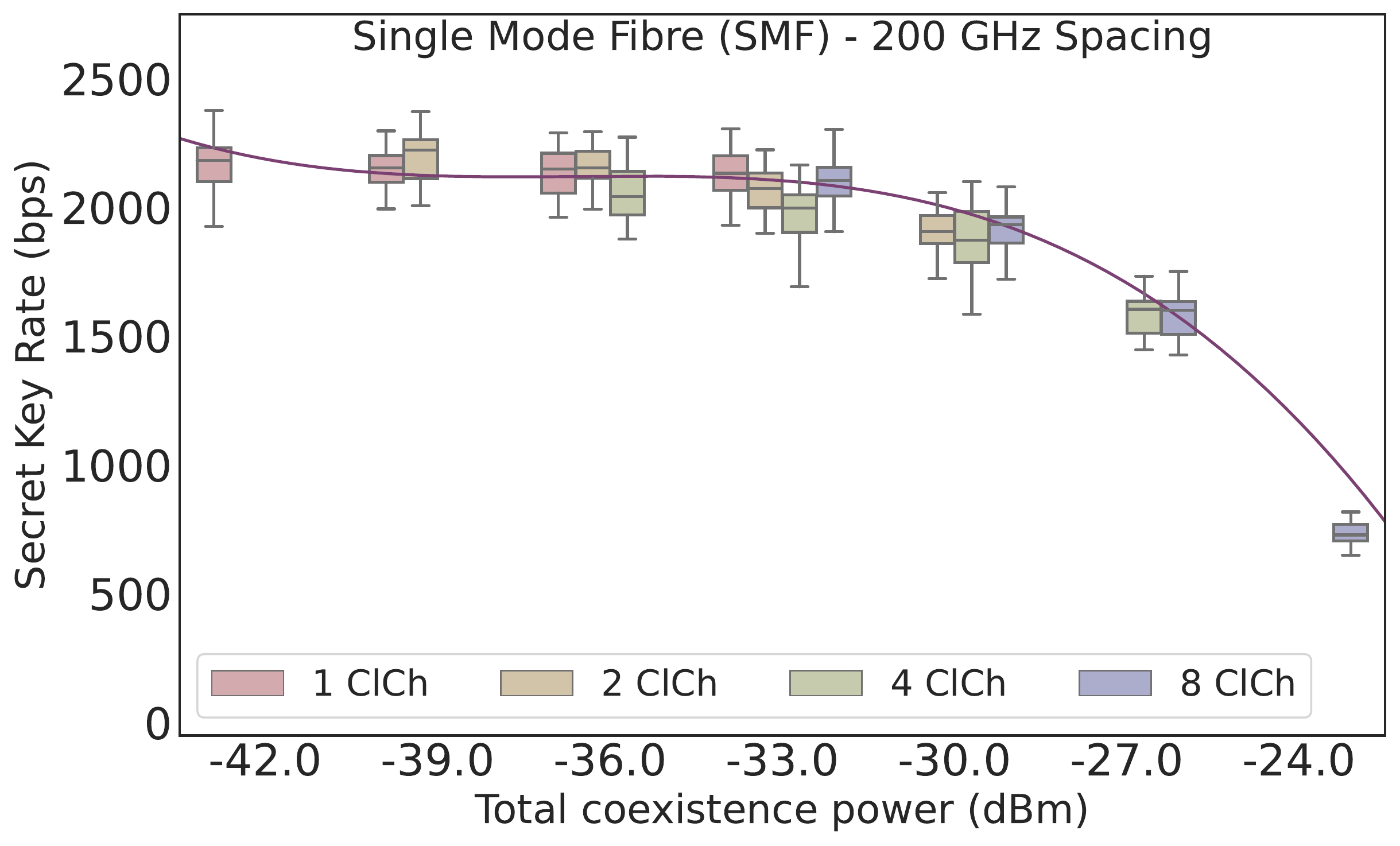} 
        \label{smf_200_skr}
        } \\
        (a) & (b) \\ 
        \subfloat { \includegraphics[width=\figureSize\linewidth]{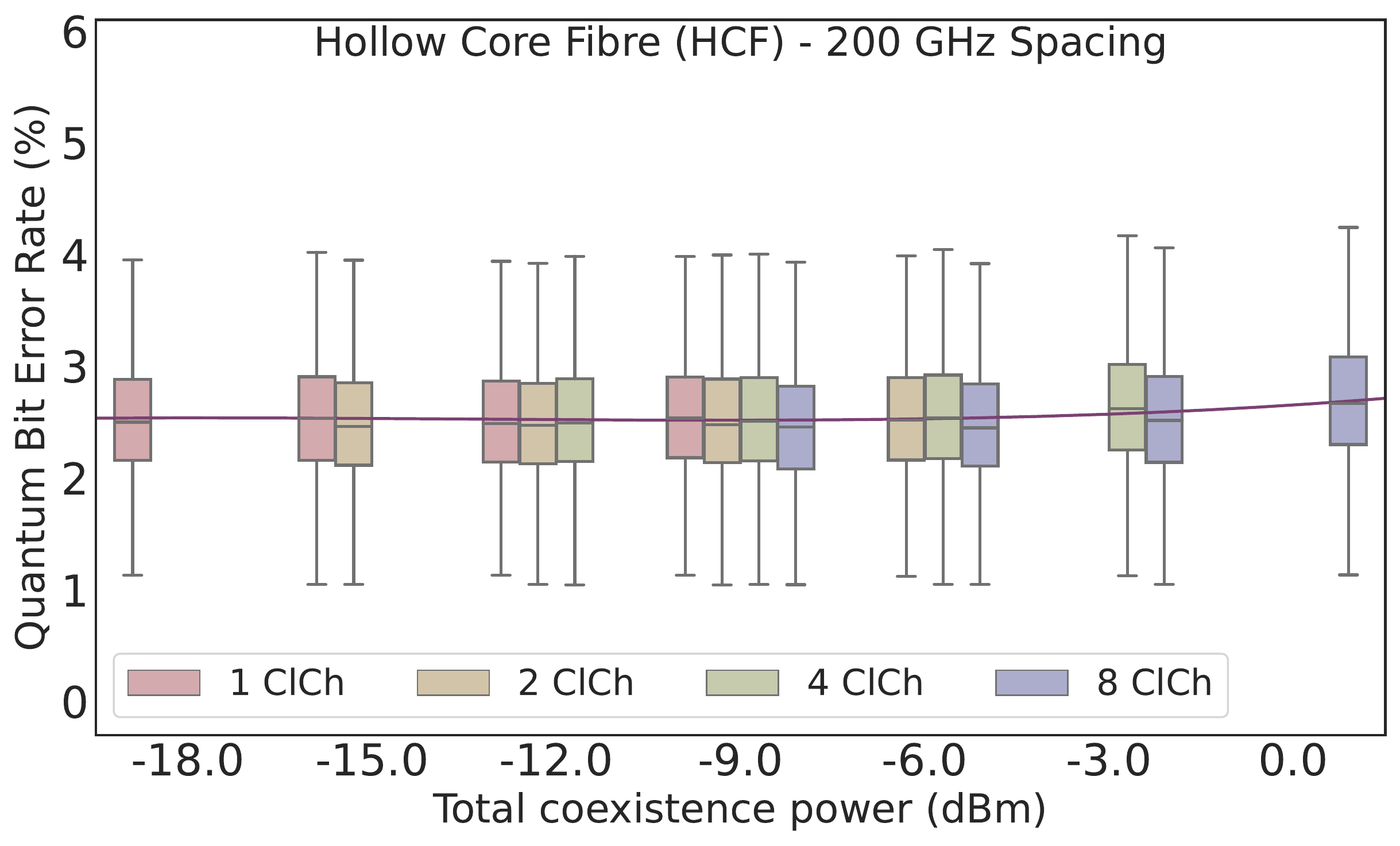}             \label{hcf_200_qber} } &
        \subfloat { \includegraphics[width=\figureSize\linewidth]{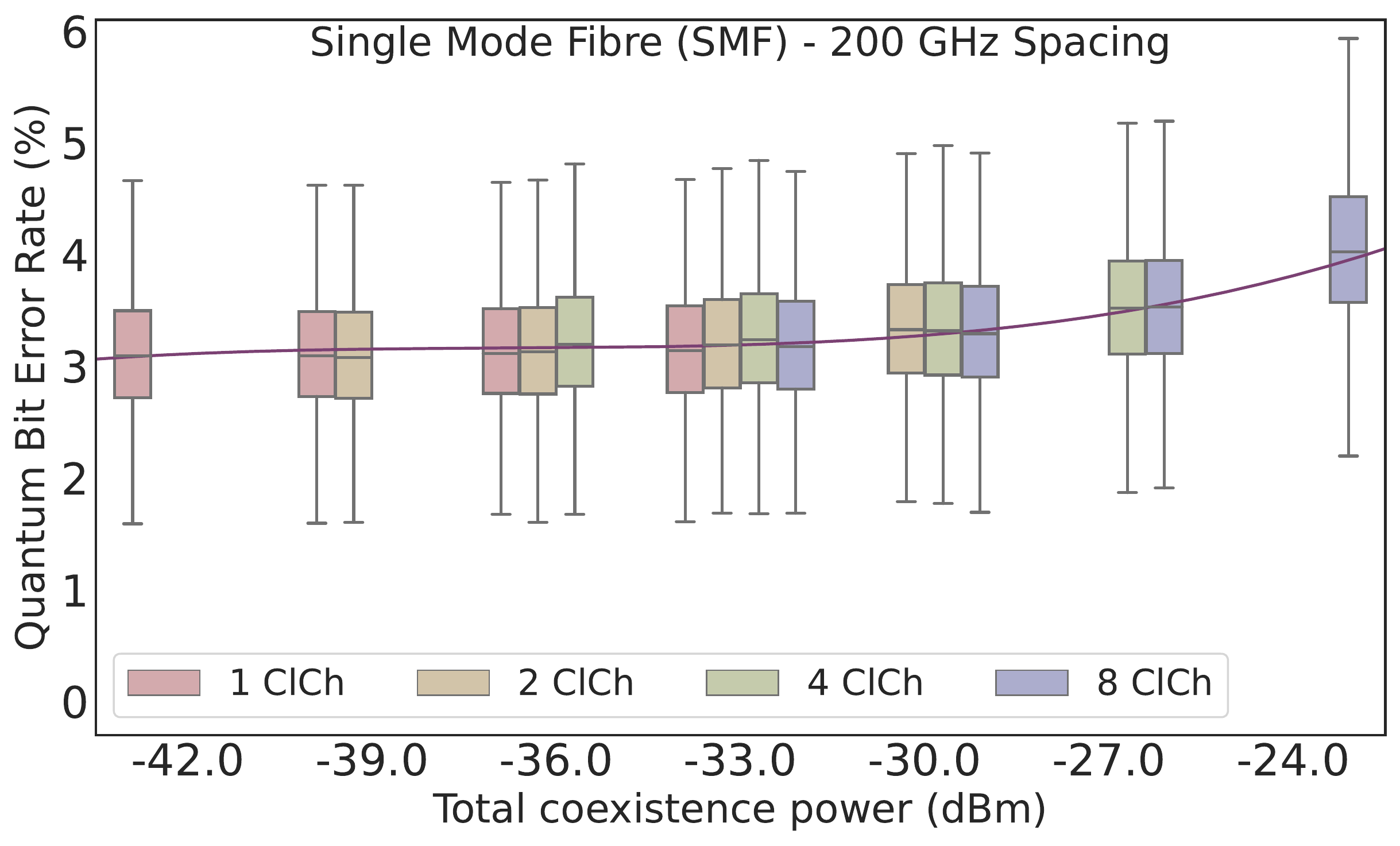}       \label{smf_200_qber} } \\
         (c) & (d) \\ 
    \end{tabular}
    \caption{ a) Average SKR versus total coexistence power using HC-NANF. b) Average SKR versus total coexistence power using SMF. c)~Average QBER versus total coexistence power using HC-NANF. d) Average QBER versus total coexistence power using SMF. 200~GHz spacing between quantum and classical channels. The solid lines in the figures are trend-line fit.}
    \label{fig:All_Charts_200}
    \setcounter{figure}{3}
\end{figure*}
}

\subsection{Co-propagation of Quantum and Classical Channel Coexistence without Fibre - System Characterisation}

To investigate the performance of the filtering stages in our system, we implemented the best-case scenario using the same coexistence scheme in \Cref{fig:testbed} without using fibres; hence removing most of the nonlinearites in the system. Since no fibres are used, the degradation in the quantum channel performance is mainly due to the photon leakage to the Bob-QKD unit from filtering deficiencies. As shown in \Cref{fig:syschar}, the average SKR stays within the expected range between 2100-2300~bps for a total coexistence power of -12~dBm which is higher than the maximum total coexistence power of -24~dBm used in SMF verifying that the degradation in the SMF fibre case is mostly due to nonlinear effects i.e. Raman scattering. This comparison is to demonstrate the effectiveness of the filtering stages in preventing photon leakage to the Bob QKD-unit for the power levels which were used in SMF. Moreover, the 1,2,4 and 8 channels symbols shown in \Cref{fig:syschar}, corresponds to the number of channels and their aggregated power. For example, 2 channels at -9~dBm corresponds to a power of -12~dBm per channel, 4 channels at -9~dBm corresponds to a power of -15~dBm per channel, and finally 8 channels at -9~dBm corresponds to a power of -18~dBm per channel.

\label{sec:Characterisation}
\begin{figure}[h]
    \centering
    \includegraphics[width=1\linewidth,clip]{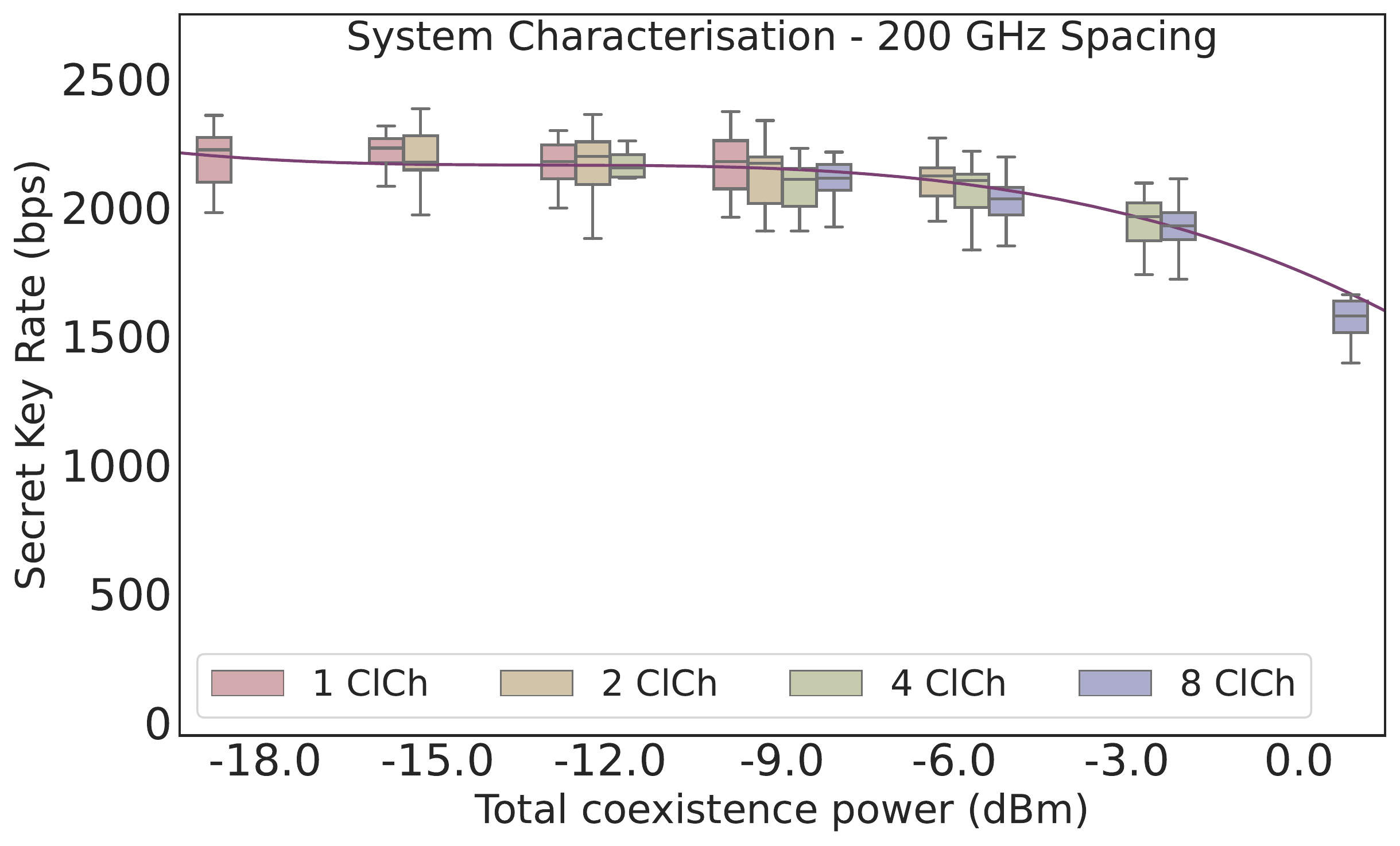}
    \caption{Average SKR versus total coexistence power without fibre - system characterisation at 200~GHz spacing.}
    \label{fig:syschar}
    \setcounter{figure}{5}
\end{figure}

\subsection{Co-propagation of Quantum and Classical Channel Coexistence over HC-NANF and SMF with different Coexistence Power and Channels Spacing}
\label{sec:Co-propagation_results}

{\newcommand\figureSize{0.475}
\begin{figure*}[!t]
    \centering
    \begin{tabular}{cc}
        \subfloat { \includegraphics[width=\figureSize\linewidth]{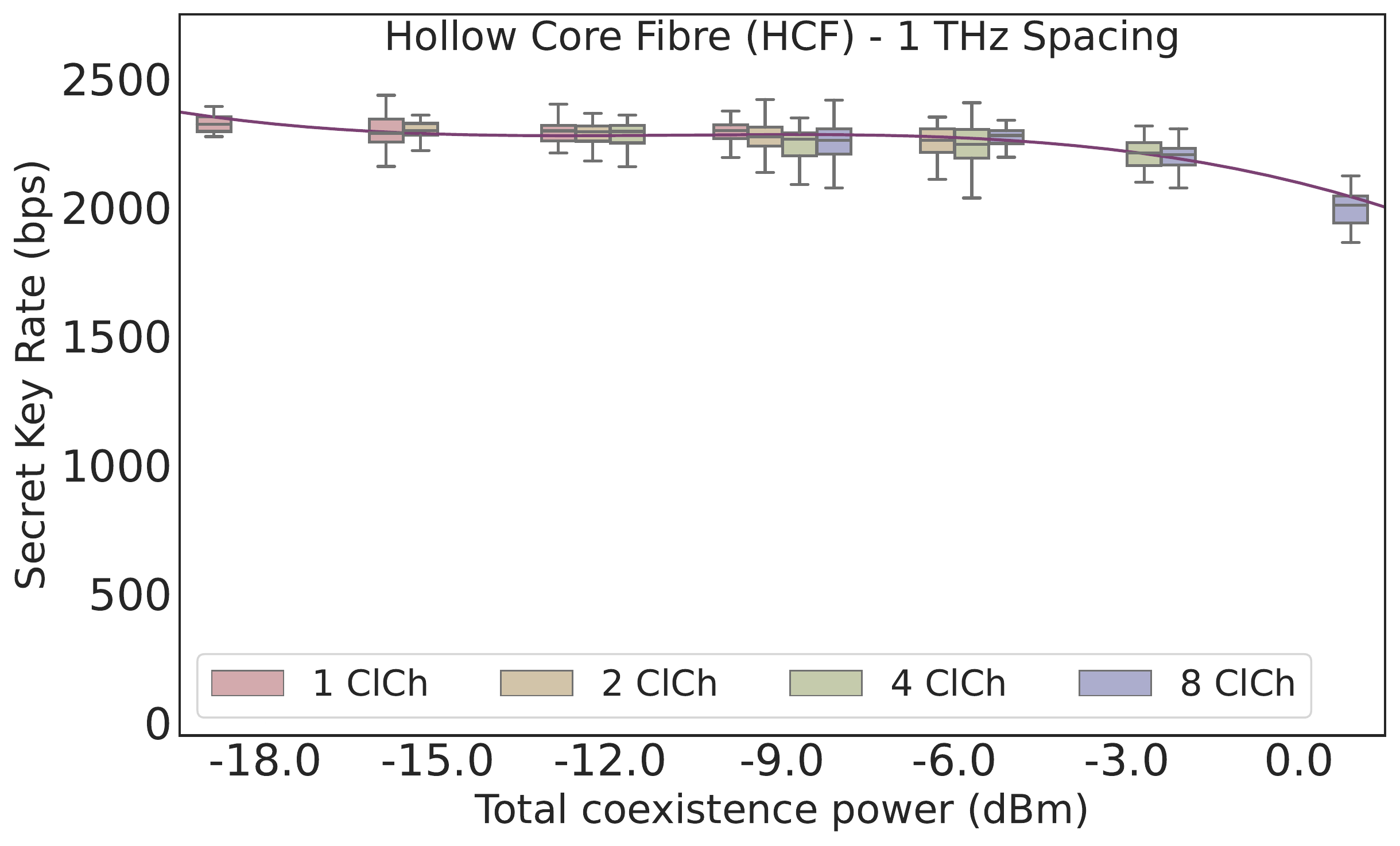}             \label{hcf_1000_skr} } & 
        \subfloat { \includegraphics[width=\figureSize\linewidth]{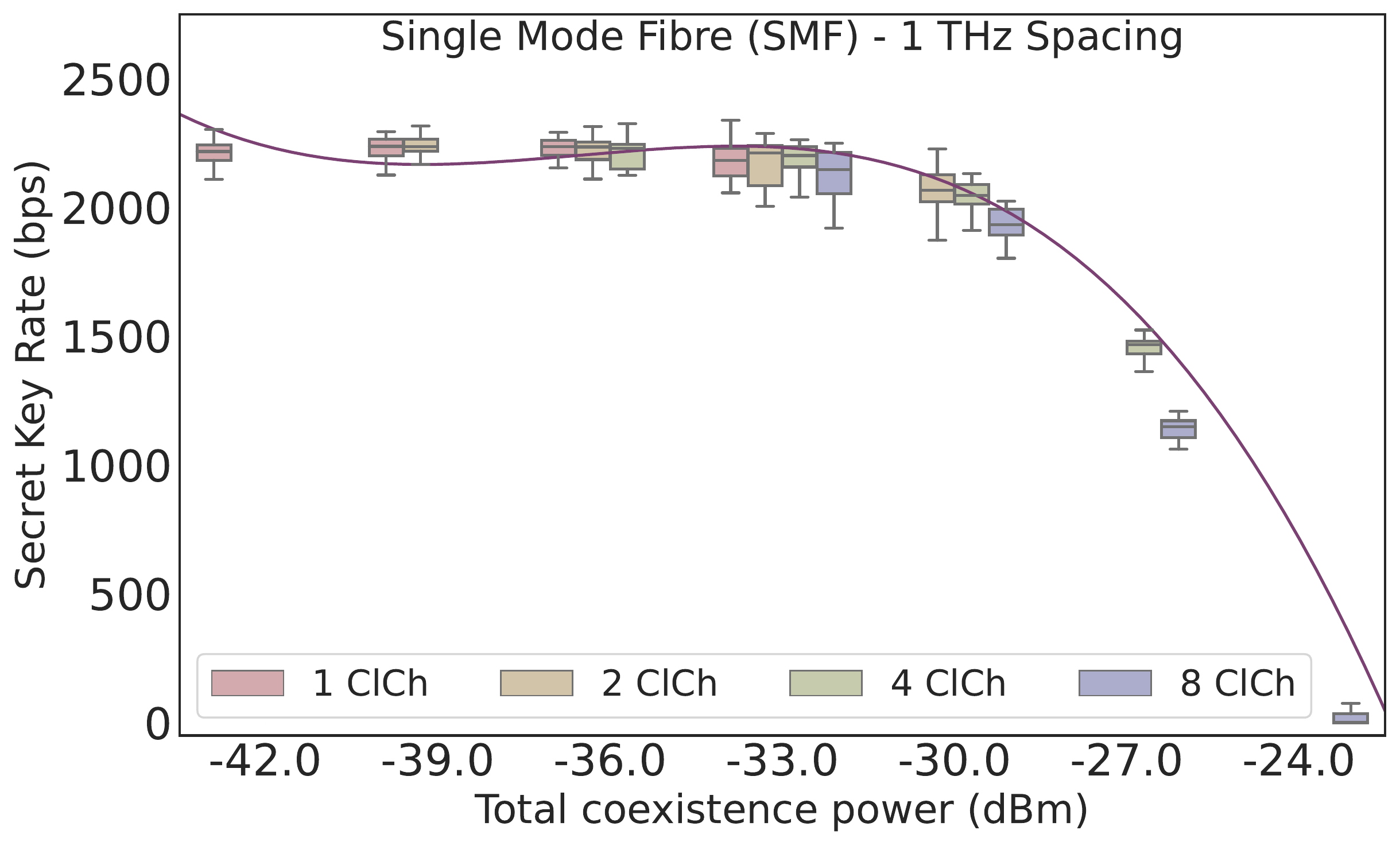} 
        \label{smf_1000_skr}
        } \\
        (a) & (b) \\ 
        \subfloat { \includegraphics[width=\figureSize\linewidth]{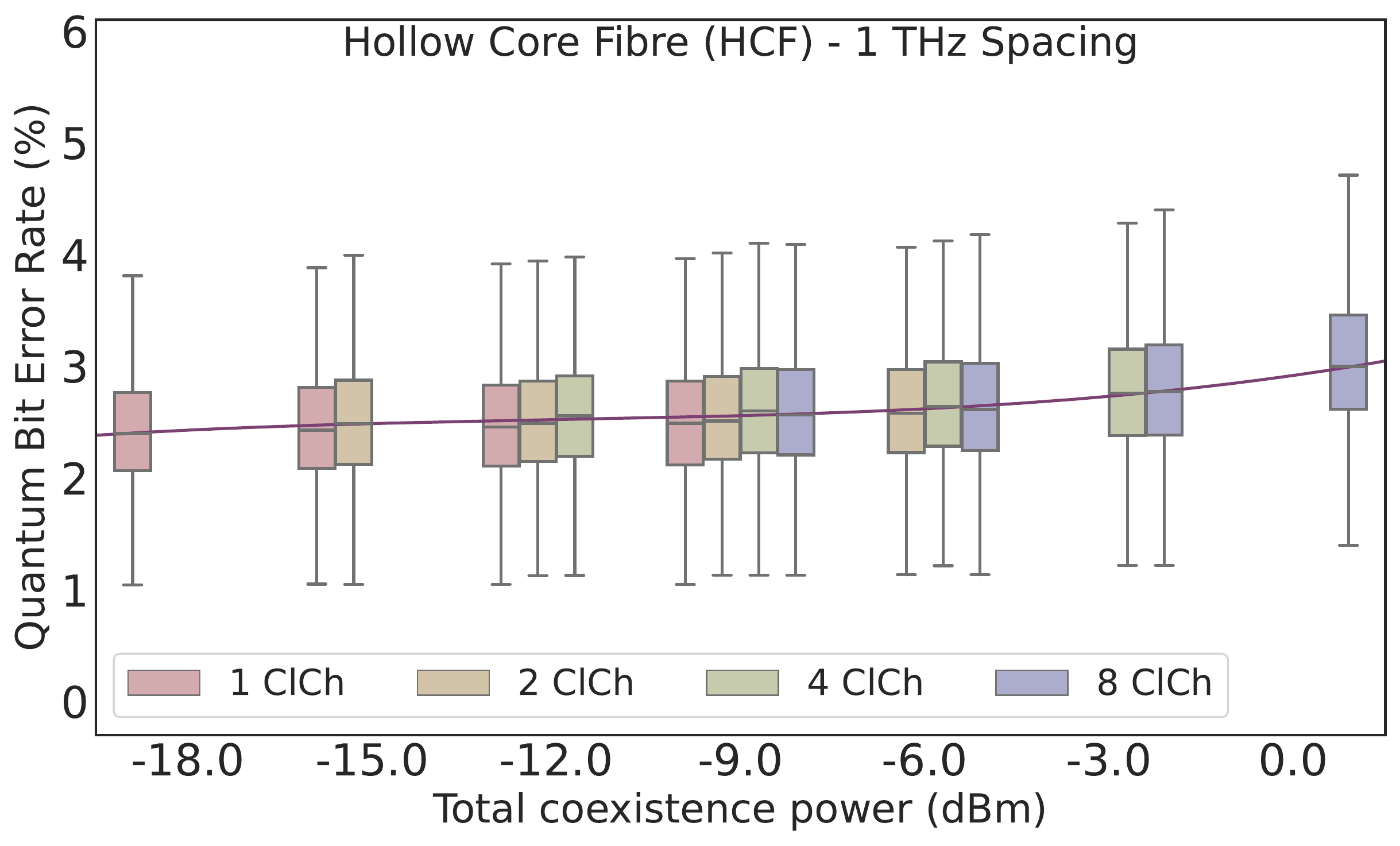}             \label{hcf_1000_qber} } &
        \subfloat { \includegraphics[width=\figureSize\linewidth]{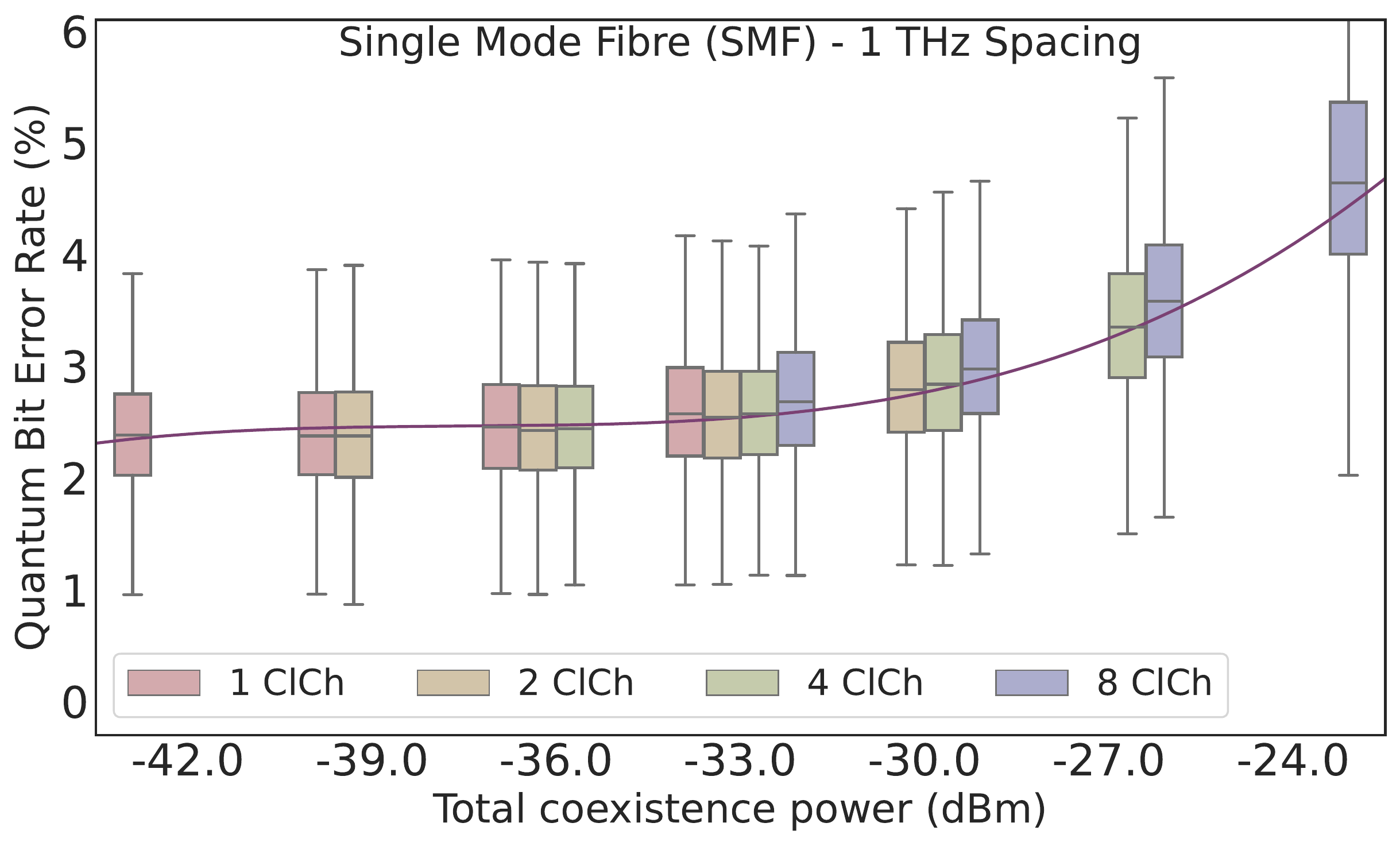}       \label{smf_1000_qber} } \\
         (c) & (d) \\ 
    \end{tabular}
    \caption{a) Average SKR versus total coexistence power using HC-NANF. b) Average SKR versus total coexistence power using SMF. c)~Average QBER versus total coexistence power using HC-NANF. d) Average QBER versus total coexistence power using SMF. 1~THz spacing between quantum and classical channels. The solid lines in the figures are trend-line fit.}
    \label{fig:All_Charts_1000}
\end{figure*}
}

\Cref{fig:All_Charts_200} and \Cref{fig:All_Charts_1000} show the experimental evaluation of SKR and QBER of the quantum channel with different spectral spacing between the classical and quantum channels based on the calculation in Section \ref{sec:simulation}. For a spacing of 1.6~nm (200~GHz) in SMF (the best case scenarios for SMF considering both Raman scaterring and FWM), \Cref{smf_200_skr}  and \Cref{smf_200_qber} show that the SKR value drops from $\approx$ 2300 bps to $\approx$ 600 bps while the QBER values increase from $\approx$ 3\% to $\approx$ 4.2\% when coexisting 8 classical channels at a total coexistence power of -24~dBm. This significant 73\% drop in the SKR and 40\% rise in the QBER is due to high optical nonlinearity in SMF causing a high noise leakage to the Bob-QKD unit. Using the same spacing of 1.6~nm in HC-NANF, \Cref{hcf_200_skr} and \Cref{hcf_200_qber} show that, the SKR and QBER values without the presence of any classical channel (no coexistence) are similar to the values when coexisting 8 classical channels at a total coexistence power of 0~dBm. This is due to the ultra-low optical nonlinearity in HC-NANF. 
For a spacing of 8~nm (1~THz) in SMF (the worst case scenarios for SMF considering Raman scaterring) and using the same coexistence power of -24~dBm, the SKR drops to zero as the QBER exceeds the operational threshold of 5.2\% of the QKD as shown in \Cref{smf_1000_skr} and \Cref{smf_1000_qber}. This is because the 8~nm spacing between the quantum and classical channels resulted in the peak of the Raman scattering spectrum to be located within the bandwidth of the BPF filter used to filter the quantum channel; hence the system stops generating keys due to excessive Raman noise leakage over the quantum channel. With the same spacing of 8~nm and coexistence power of 0~dBm in HC-NANF, \Cref{hcf_1000_skr} and \Cref{hcf_1000_qber} show that, the SKR drops from $\approx$ 2300 bps to $\approx$ 2050 bps while the QBER values increase from $\approx$ 2.2\% to $\approx$ 3.1\% using HC-NANF.

In both scenarios, the back-to-back losses of the quantum channel is similar when using both SMF and HC-NANF (10.5~dB) and the highest coexistence power (0~dBm) in the HC-NANF is 250~times higher than the highest coexistence power (-24~dBm) in SMF. Even though the coexistence power is much higher is HC-NANF, the ultra-low nonlinear effects of the due to its hollow core preserve or have a slight effect on the SKR and QBER of the quantum channel. This proves the suitability of HC-NANF as an excellent transmission medium for high power coexistence of quantum and classical channels. 


\section{Conclusion}
\label{sec:conclusion}

The coexistence of a DV-QKD channel and 8 × 200 Gbps 16-QAM classical channels was successfully demonstrated over a 2-km long Hollow Core Nested Antiresonant Nodeless Fibre for a record-high coexistence transmission of 1.6 Tbps. The best and worst spectral positions of classical channels were determined via numerical calculation of the impact of the nonlinear effects such as Raman scattering and four-wave mixing on the quantum channel performance. In the best case scenario with 1.6~nm spacing between the quantum and classical channels (200~GHz) at -24~dBm total coexistence power in SMF, the SKR dropped by 73\%, whereas, at 0~dBm total coexistence power in HC-NANF (250 times higher than the power used in SMF), the SKR was preserved. In the worst case scenario 8~nm spacing between the quantum and classical channels (1~THz - Raman spectrum peak at the quantum channel wavelength) and using the same powers, the SKR dropped to 0\% in SMF meaning no keys were generated, whereas in HC-NANF the SKR dropped only by 10\%. This significant difference in the QKD performance proves the advantage of using HC-NANF to provide a seamless coexistence of quantum and classical channels for future quantum networks with a minimal effect on the quantum channel performance regardless of the number and power of the classical channels.

\section*{Acknowledgement}
This work was funded by EU funded project UNIQORN (820474) and the EPSRC Airguide Photonics Collaboration Fund Award (ref: 517129) (EP/P030181/1). Part of the research leading to this work has been supported by the Quantum Communication Hub funded by the EPSRC grant ref. EP/T001011/1 and the ERC LightPipe project (grant n$^{\circ}$ 682724).

\bibliographystyle{IEEEtran}
\bibliography{IEEEabrv,refs}









\end{document}